\documentclass[cameraready]{Interspeech}

\usepackage{float}
\usepackage{flushend}
\usepackage{bm}
\usepackage{subcaption}

\title{Room Impulse Response Completion Using Signal-Prediction Diffusion Models Conditioned on Simulated Early Reflections}

\author[affiliation={1}, orcid=0000-0002-4158-6218, correspondingauthor]{Zeyu}{Xu}
\author[affiliation={2}, orcid=0000-0002-6051-6346]{Andreas}{Brendel}
\author[affiliation={1}, orcid=0000-0003-3491-3644,]{Albert G.}{Prinn}
\author[affiliation={1}, orcid=0000-0002-2613-8046,]{Emanu\"{e}l A. P.}{Habets}

\address{
    $^1$ International Audio Laboratories Erlangen$^\dagger$, Germany \\
    $^2$ Fraunhofer Institute for Integrated Circuits (IIS), Erlangen, Germany \\
    $^{\dagger}$A joint institution of the Friedrich-Alexander-Universität Erlangen-Nürnberg (FAU) \\ and Fraunhofer Institute for Integrated Circuits (IIS)
}

\email{zeyu.xu@audiolabs-erlangen.de}

\usepackage{comment}
\usepackage[sort]{cite}
\usepackage{booktabs}
\usepackage{adjustbox}
\usepackage{multirow}
\usepackage{graphicx}
\usepackage{subcaption}
\newcommand{\syscell}[2]{%
  \begin{adjustbox}{valign=t}%
    \begin{tabular}{@{}l@{\hspace{6pt}}l@{}}#1 & #2\end{tabular}%
  \end{adjustbox}%
}

\begin{document}

\maketitle

\begin{abstract}
Room impulse responses (RIRs) are fundamental to audio data augmentation, acoustic signal processing, and immersive audio rendering. While geometric simulators such as the image source method (ISM) can efficiently generate early reflections, they lack the realism of measured RIRs due to missing acoustic wave effects. We propose a diffusion-based RIR completion method using signal-prediction conditioned on ISM-simulated direct-path and early reflections. Unlike state-of-the-art methods, our approach imposes no fixed duration constraint on the input early reflections. We further incorporate classifier-free guidance to steer generation toward a target distribution learned from physically realistic RIRs simulated with the Treble SDK. Objective evaluation demonstrates that the proposed method outperforms a state-of-the-art baseline in early RIR completion and energy decay curve reconstruction.
\end{abstract}

\section{Introduction}

The generation of room impulse responses, RIRs, is central to immersive spatial audio rendering, acoustic array signal processing, and the construction of large scale machine learning datasets. Traditional physics based methods are broadly classified into geometrical and numerical wave methods. Geometrical methods, including the image source method, ISM~\cite{Allen1979,Borish1984,Xu2024}, ray tracing~\cite{Krokstad1968}, beam tracing~\cite{Funkhouser2004}, and delay networks~\cite{Jot1991}, model individual reflection paths. Numerical wave methods instead solve the discretized wave equation using finite difference time domain~\cite{Botteldooren1995}, finite element~\cite{Shuku1973}, boundary element~\cite{Bai1992}, or finite volume techniques~\cite{Bilbao2015}. Recently, deep learning approaches such as neural acoustic field models~\cite{Luo2022}, Fast RIR~\cite{Ratnarajah2022}, and physics informed neural network RIR~\cite{Karakonstantis2024} have been proposed in the literature.

More recently, RIR completion has emerged as a new task in RIR generation, aiming to complete the full RIR after a 50~ms~\cite{Lin2025} or 80 ms~\cite{Kim2024} window (RIR head), comprising the direct path and early reflections, with the direct-path delay removed. This new task assumes that the direct path and early reflections contain sufficient information about the room, i.e., its geometry and wall materials, to predict the late part of the RIR. However, existing state-of-the-art methods for RIR completion~\cite{Kim2024,Lin2025} typically require an RIR-head input truncated from a fully simulated or measured RIR, whereas many geometric RIR simulation implementations, e.g., the ISM~\cite{Habets2006,Scheibler2018}, accept the maximum reflection order as an input parameter. This is a limitation of existing methods, as using an incomplete RIR head from a low-order ISM simulation would lead to discontinuities in the completed RIR.


Within generative modeling, diffusion models~\cite{ho2020denoising} offer a powerful framework for learning complex data distributions and have recently been applied to RIR tasks. Examples include DiffusionRIR~\cite{Della2025}, which interpolates RIRs at unseen positions, and Gencho~\cite{Lin2026}, which estimates RIRs from reverberant speech. A central design choice in diffusion models is the target predicted at each diffusion step. The original DDPM formulation~\cite{ho2020denoising} introduced noise prediction, which remains popular due to its stable training dynamics. However, equivalent parameterizations exist, including velocity prediction ($\mathbf{v}$-prediction), as adopted in Gencho~\cite{Lin2026}. Despite these advances, current diffusion-based RIR approaches focus on interpolation or full RIR estimation, and RIR completion methods typically rely on complete early-reflection information. To date, there is no diffusion framework tailored to RIR completion from incomplete early responses, nor a systematic investigation of applying signal prediction ($\mathbf{x}$-prediction) as the target in diffusion models for generating physically coherent RIRs without discontinuities under partial conditioning.


This paper proposes an RIR completion method based on an $\mathbf{x}$-prediction diffusion model conditioned solely on simulated direct-path and early reflections obtained from the ISM. The proposed approach generates a complete full-band RIR even when the RIR head is incomplete, whereas existing methods may introduce temporal discontinuities. In particular, it does not require access to the truncated early reflections of a fully simulated or measured RIR. 
To more accurately assess the performance of the proposed method under realistic conditions, two paired datasets with identical room configurations are constructed for classifier-free guidance (CFG) training: one consisting solely of ISM simulations and the other comprising RIRs simulated using a numerical wave method. 
When trained with CFG, the model produces more realistic RIRs than a baseline diffusion model, even when the training data are dominated by ISM-simulated RIRs that lack wave interference effects and include only a small proportion of realistic examples. Furthermore, incorporating an energy decay curve (EDC) loss~\cite{Mezza2025} improves the preservation of physically consistent energy decay characteristics.

\section{Proposed Method}

\subsection{Conditioned Diffusion Models with $\mathbf{x}$-Prediction}
We define the noise-free target RIR of length $K$ as $\mathbf{x}_0 \in \mathbb{R}^{K}$, and noisy RIR at time step $t$ in the forward and backward diffusion as $\mathbf{x}_t \in \mathbb{R}^{K}$. The forward diffusion process for $t = 1,\dots,T$ can be written as 
\begin{equation}
    \mathbf{x}_t = \sqrt{\bar{\alpha}_t} \mathbf{x}_0 + \sqrt{1-\bar{\alpha}_t} \bm{\epsilon}_t~, \label{eq:forward_diffusion}
\end{equation}
where $\bm{\epsilon}_t\sim\mathcal N(\mathbf{0},\mathbf{I})$ is the added standard Gaussian noise in $\mathbb{R}^{K}$, and $\bar{\alpha}_t = \prod_{t^\prime=1}^t \alpha_{t^\prime}$ is the accumulated signal retention after $t$ time steps and $\alpha_{t^\prime}$ is the signal retention at time step $t^\prime$. They are obtained using the cosine schedule~\cite{nichol2021improved}. 

We aim to reconstruct the target RIR $\mathbf{x}_0$ using a result of a low-order ISM simulation $\mathbf{c} \in \mathbb{R}^{K}$ as the conditioner of the diffusion model. For $\mathbf{x}$-prediction, the neural network $\mathcal{X}_\theta(\cdot)$ parameterized by $\theta$ is trained to directly predict the target RIR $\mathbf{x}_0$ in \eqref{eq:forward_diffusion}, i.e., $\hat{\mathbf{x}}_{0}= \mathcal{X}_{\theta}(\mathbf{x}_t,\mathbf{c},t) \approx \mathbf{x}_0$.
%
During inference, the same noise schedule as in training is employed, and the update $\mathbf{x}_{t-1} = \bm{\mu}_\theta(\mathbf{x}_t,t) + \sigma_t \bm{\epsilon}_t$ is applied recursively.
The reverse mean $\bm{\mu}_\theta(\mathbf{x}_t,t)$ is calculated using 
\begin{equation}
    \bm{\mu}_\theta(\mathbf{x}_t,t) = \frac{\sqrt{\bar\alpha_{t-1}}(1 - \alpha_t)}{1-\bar\alpha_t} \hat{\mathbf{x}}_{0}+ \frac{\sqrt{\alpha_t}\bigl(1-\bar\alpha_{t-1}\bigr)}{1-\bar\alpha_t} \mathbf{x}_t~,
\end{equation}
where the standard deviation is $\sigma_t = \sqrt{\frac{(1-\bar{\alpha}_{t-1})(1 - \alpha_t)}{1-\bar{\alpha}_t}}$, and the noise $\bm{\epsilon}_t$ is not added in the last step. 

\subsection{Network Structure}
We use the 1D U-Net architecture~\cite{ho2020denoising} to predict the target RIR $\mathbf{x}_0$ from the noisy sample $\mathbf{x}_t$ within the diffusion process. The ISM-based conditioning signal $\mathbf{c}$ is concatenated with $\mathbf{x}_t$ along the channel dimension, yielding the input $[\mathbf{c}, \mathbf{x}_t] \in \mathbb{R}^{K\times 2}$. 
Two major modifications are made to the 1D U-Net proposed in~\cite{ho2020denoising}: i)~The required downsampling factor of the U-Net encoder depends on the maximum RIR length and the target bottleneck length. In our model, $7$~stride-2~downsamples are applied, supporting up to $32\,768$ RIR samples and a bottleneck length of $256$. ii)~In the bottleneck, a 6-layer residual dilated Conv1D stack with dilations $1,2,4,8,16,32$ is applied to capture long-range temporal structure, such as late reverberation. 

\subsection{Loss Functions}
For the model output $\hat{\mathbf{x}}_{0}=[\hat{x}_0[0], \hat{x}_0[1],\dots,\hat{x}_0[K-1]]^T$ and the target $\mathbf{x}_{0}=[x_0[0], x_0[1],\dots,x_0[K-1]]^T$, the MSE loss is given by
\begin{equation} \label{eqn:mse}
    \mathcal{L}_{\textrm{MSE}} = \frac{1}{K} \sum_{n=0}^{K-1} \left(x_{0}[n] - \hat{x}_{0}[n]\right)^2.
\end{equation}
The EDC of an RIR $\mathbf{x}_0$ can be calculated using a discrete version of Schroeder's backward integration~\cite{schroeder1965}: $E[n]  = \sum_{k=n}^{K-1} x_0^2[k] $ for $n = 0,1,\dots,K-1$.
The normalized EDC in logarithmic scale is given by $E_{\textrm{dB}}[n]  = 10 \log_{10} ( E[n]/E[0] )$.
The EDC is further clamped by setting $E_{\textrm{dB}}[n] = \max(E_{\textrm{dB}}[n], E_{\textrm{min}})$ with $E_{\textrm{min}} = -60$ dB. With the EDCs of the predicted and target RIRs, given by $\widehat{E}_{\textrm{dB}}$ and $E_{\textrm{dB}}$, respectively, we calculate the EDC loss as
\begin{equation}
    \mathcal{L}_{\textrm{EDC}} = \sum_{n=0}^{K-1} w[n]\left|\widehat{E}_{\textrm{dB}}[n] - E_{\textrm{dB}}[n] \right|, \label{eq:EDC_batch_i}
\end{equation}
where $w[n]$ forms a normalized uniform weighting over time samples for which $E_{\textrm{dB}}[n] \geq -60$.

The total loss is a weighted combination of the MSE and EDC loss
\begin{equation}
    \mathcal{L}_{\textrm{total}} =  \mathcal{L}_{\textrm{MSE}} + \lambda  \mathcal{L}_{\textrm{EDC}} ~,\label{eq:total_loss}
\end{equation}
where $\lambda \geq 0$ is a scaling factor.

\subsection{Classifier-Free Guidance}\label{sec:CFG}
For CFG~\cite{Ho2022}, we train a single denoising network to operate in both conditional and unconditional mode by randomly replacing the conditioner with an all-zero vector $\mathbf{c}=\mathbf{0}\in \mathbb{R}^K$ with probability $p_{\textrm{CFG}}$. This conditioning dropout forces the model to jointly learn the conditional and unconditional RIR distribution guided by the early reflections.


At inference, we apply CFG by evaluating the denoiser twice per reverse step, once with the conditioning signal and once with the null condition. With the two predictions in each reverse diffusion step, $\hat{\mathbf{x}}_{0}^{\textrm{c}} = \mathcal{X}_\theta(\mathbf{x}_t,\mathbf{c},t)$ and $\hat{\mathbf{x}}_{0}^\textrm{uc} = \mathcal{X}_\theta(\mathbf{x}_t,\mathbf{0},t)$, the combined prediction is
\begin{equation}
    \hat{\mathbf{x}}_{0}^\textrm{CFG} = \hat{\mathbf{x}}_{0}^\textrm{uc
    } + s \, (\hat{\mathbf{x}}_{0}^\textrm{c} - \hat{\mathbf{x}}_{0}^\textrm{uc})~, \label{eq:x0_CFG_prediction}
\end{equation}
where $s\geq1$ is the guidance scale that controls the tradeoff between condition adherence and sample diversity.

\section{Experimental Setup}
\subsection{Datasets}

Two paired RIR datasets were generated for $25$ shoebox rooms with randomly sampled dimensions under consistent simulation conditions using pyroomacoustics~\cite{Scheibler2018} and the Treble SDK~\cite{TrebleSDKDocs}. Floor materials were randomly selected from carpet types, while the ceiling and four walls were randomly assigned gypsum-based materials. For the Treble SDK dataset only, one upholstered sofa, two wooden chairs, and one wooden table were randomly placed on the floor, with identical furniture materials across rooms. Each room contained $10$ point sources and $40$ omnidirectional receivers, all positioned randomly, subject to a minimum distance from the walls and furniture.

The dataset generated using only the ISM in pyroomacoustics is referred to as the \emph{ISM dataset}, whereas the dataset generated using both geometric and numerical wave simulation in the Treble SDK is referred to as the \emph{Treble dataset}. It has recently been shown that RIRs produced with the numerical wave simulator in the Treble SDK enable learning-based acoustic signal processing methods to achieve performance comparable to that obtained with measured RIRs~\cite{Gotz2025}. Importantly, room geometry, wall absorption coefficients, and source and receiver positions are identical across the two datasets. The Treble dataset additionally captures wave interference due to furniture as well as diffraction at edges and corners.

Each dataset comprised $10\,000$ RIRs and was randomly split in an $8:1:1$ ratio into training, validation, and test sets. All RIRs were normalized to unit peak amplitude. Direct-path delays were removed from both RIRs and conditioning signals, except for a $2.5$~ms segment retained to preserve the first arrival. RIRs were truncated or zero-padded to $K = 24\,576$ samples, corresponding to $1.536$~s at a sampling rate of $16$~kHz.

Since the original Echo2Reverb implementation performs RIR completion up to $2.5$~s at $48$~kHz, all signals were first upsampled from $16$~kHz to $48$~kHz and zero-padded to match the required length. After inference, the generated outputs were low-pass filtered, downsampled, and truncated to $K = 24\,576$ samples for evaluation.

\subsection{Conditioner Configurations}\label{sec:conditioner_study}



\begin{figure}[!t]
    \centering
    \includegraphics[width=0.85\linewidth]{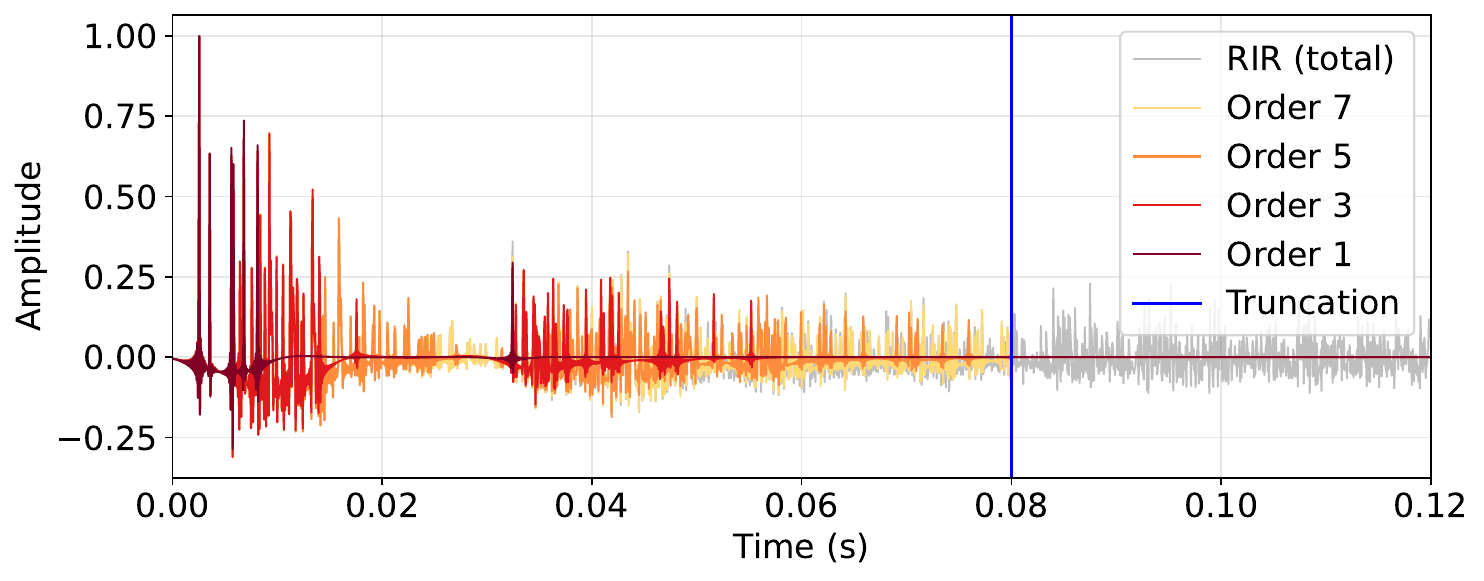}
    \vspace{-3mm}
    \caption{Examples of ISM conditioners with different maximum reflection orders $1,3,5,7$ and the full RIR. The conditioners are truncated to $80$~ms such that they can be used as inputs to both the proposed model and Echo2Reverb.} \vspace{-1em}
    \label{fig:conditioners}
\end{figure}

The baseline model Echo2Reverb~\cite{Kim2024} used the first $80$ ms of a full RIR as conditioning input to generate late reverberation. In practice, however, early reflections simulated with the ISM can be obtained more efficiently by specifying a maximum reflection order rather than a fixed time window.
Accordingly, ISM-based conditioners $\mathbf{c}$ of length $80$~ms were generated with maximum reflection orders of $1$, $3$, $5$, and $7$. To remain consistent with the original baseline configuration, Echo2Reverb was conditioned on early reflections obtained by truncating the full RIR to $80$ ms. In addition, to evaluate its behavior under order-limited conditions, ISM conditioners with reflection orders $5$ and $7$ were also provided as inputs. The ISM conditioners were generated using the same room parameters as those employed for full RIR simulation in the ISM dataset.
Examples of truncated ISM conditioners are shown in Fig.~\ref{fig:conditioners}. Notably, even a reflection order of $7$ does not fully populate the $80$~ms window. The proposed model does not require truncation of the conditioning signal.

\subsection{Experiments}\label{sec:experiments}

Two experiments were conducted to evaluate the proposed method and Echo2Reverb. In the first experiment, only the ISM dataset was used. In the second experiment, a \emph{combined dataset} was constructed by randomly selecting from the ISM and Treble datasets with a ratio of $8:2$ for training and validation. Only the Treble dataset was used for testing. 
The CFG probability $p_{\textrm{CFG}}$ was set to $0.2$ during training and validation. 
Echo2Reverb was trained and validated using the same dataset configuration; however, in contrast to the CFG-based training pipeline of the proposed method, the conditioning signals were not dropped. For the proposed method, the guidance scale was set to $1$ during inference, corresponding to fully conditioned sampling.
Across both experiments, the maximum reflection order of the ISM conditioner (described in Sec.~\ref{sec:conditioner_study}) was varied. Furthermore, the proposed model was trained using either the MSE loss (\ref{eqn:mse}) or the total loss (\ref{eq:total_loss}) with an empirically chosen scaling factor $\lambda=10^{-5}$.

\section{Evaluation}
\subsection{Metrics}
We evaluated reconstruction errors between the predicted RIR $\hat{\mathbf{x}}_0$ and the noise-free target RIR $\mathbf{x}_0$ for different sections of the RIRs.
The first section covers the early $80$~ms window, where the conditioner $\mathbf{c}$ may overlap with the full RIR, as shown in Fig.~\ref{fig:conditioners}. The objective is to quantify the mismatch between the predicted and target RIRs only in the residual components, i.e., between $\mathbf{r} = \mathbf{x} - \mathbf{c}$ and $\hat{\mathbf{r}} = \hat{\mathbf{x}} - \mathbf{c}$. 
Since the ISM conditioner may span the $80$~ms window differently as the maximum reflection order changes, we calculated the following residual energy ratio (RER) for the early $80$~ms
\begin{equation}
\textrm{RIR}^{\leq \textrm{80}}_{\textrm{RER}}
= \frac{\sum_{n=0}^{K_\textrm{80}-1}\hat{r}[n]^2}{\sum_{n=0}^{K_\textrm{80}-1}r[n]^2}~,
\end{equation}
where $K_\textrm{80}$ denotes the first time sample after the conditioner window. 
A value of around $1$ of the RER indicates that the residual of the completed RIR and the conditioner contains a similar amount of energy as the residual of the target RIR in the $80$~ms window. 
For the RIR completion error after $80$~ms, the root mean square error (RMSE)
\begin{equation}
    \textrm{RIR}_{\textrm{RMSE}}^{>\textrm{80}} = \sqrt{\frac{1}{K-K_\textrm{80}} \sum_{n=K_\textrm{80}}^{K-1} \left( \hat{x}[n] - x[n] \right)^2}\label{eq:rir_rmse_after}
\end{equation}
was calculated.
To measure the accuracy in terms of the EDC, the mean absolute error (MAE) was calculated between the target EDC $E_{\textrm{dB}}[n]$ and the predicted EDC $\widehat{E}_{\textrm{dB}}[n]$ for the time samples where $E_{\textrm{dB}}[n]\geq -60~\textrm{dB}$
\begin{equation}
    \textrm{EDC}_{\textrm{MAE}} = \sum_{n=0}^{K-1} w[n] \left| \widehat{E}_{\textrm{dB}}[n] - E_{\textrm{dB}}[n] \right|~,\label{eq:edc_mae}
\end{equation}
where $w[n]$ is the same as used in \eqref{eq:EDC_batch_i}.
Note that all three error metrics are reported in dB.

\subsection{Experiment 1: ISM RIRs}
The objective evaluation over the ISM dataset is illustrated as Exp.~1 in Tab.~\ref{tab:evaluation}, where the mean and standard deviation of the three metrics are given for the proposed and baseline models. 

For the proposed model, the three metrics do not vary significantly as the maximum reflection order of the input ISM conditioner changes. However, including the EDC loss \eqref{eq:total_loss} lowers $\textrm{EDC}_{\textrm{MAE}}$ significantly relative to using only the MSE loss in training. When the ISM conditioner is of order 5, the best EDC errors in our model are slightly worse than those of the Echo2Reverb model fed by the same ISM conditioner. When the ISM conditioner's order increases to $7$, the early RER error $\textrm{RIR}^{\leq \textrm{80}}_{\textrm{RER}}$ is lower than for the Echo2Reverb with the same conditioner when the hybrid loss is used in our method. 


\begin{table*}[t]
\caption{Performance comparison between our proposed model and the baseline model Echo2Reverb for the two experiments. M denotes that only the MSE loss was used. M+E indicates that both the EDC and the MSE losses were used. The numbers $1,3,5,7$ and Full denote the maximum reflection order in the ISM conditioner, used as input to the trained model. For each entry, mean (standard deviation) values are shown. Arrow $\downarrow$ indicates lower values are better, and $0$ indicates that values closer to $0$ are better. The best value in each column is highlighted for Echo2Reverb and our method. The RIRs of the ISM dataset are also compared against the metrics in Exp.~2 to demonstrate the intrinsic dataset mismatch between the ISM and Treble datasets, despite equal room configurations.} 
\centering
\renewcommand{\arraystretch}{0.8}
\setlength{\tabcolsep}{3pt}
 \vspace{-1mm}
\begin{adjustbox}{max width=0.9 \textwidth}
\begin{tabular}{ll ccc ccc}
\toprule
\multicolumn{2}{c}{Source} &
\multicolumn{3}{c}{Exp.~1: ISM dataset} &
\multicolumn{3}{c}{Exp.~2: CFG training with the combined dataset} \\
\cmidrule(lr){1-2}\cmidrule(lr){3-5}\cmidrule(lr){6-8}
Method & Order
& $\textrm{RIR}^{\leq \textrm{80}}_{\textrm{RER}}\,(\textrm{dB},\,$0$)$
& $\textrm{RIR}_{\textrm{RMSE}}^{>\textrm{80}}\,(\textrm{dB},\,\downarrow)$
& $\textrm{EDC}_{\textrm{MAE}}\,(\textrm{dB},\,\downarrow)$
& $\textrm{RIR}^{\leq \textrm{80}}_{\textrm{RER}}\,(\textrm{dB},\,$0$)$
& $\textrm{RIR}_{\textrm{RMSE}}^{>\textrm{80}}\,(\textrm{dB},\,\downarrow)$
& $\textrm{EDC}_{\textrm{MAE}}\,(\textrm{dB},\,\downarrow)$ \\
\midrule

\syscell{Echo2Reverb}{} & 5   & -3.73 (6.11) & -21.09 (6.11) & \textbf{-1.01 (3.36)} & -12.21 (1.96) & -21.95 (5.35) & \textbf{9.46 (3.75)} \\
\syscell{Echo2Reverb}{} & 7   & \textbf{2.57 (7.09)} & -21.14 (6.14) & -0.35 (3.73) & -12.18 (1.83) & -22.06 (5.41) & 9.62 (3.56) \\
\syscell{Echo2Reverb}{} & Full & -- & \textbf{-21.18 (6.15)} & 0.73 (4.24) & \textbf{-12.06 (1.72)}  & \textbf{-22.07 (5.47)} & 9.62 (3.53) \\
\midrule

\syscell{Proposed}{M}   & 1 & -1.54 (1.54) & -21.63 (5.82) & 9.54 (2.19) & -1.94 (2.14) & \textbf{-23.92 (4.80)} & 10.23 (1.53) \\
\syscell{Proposed}{M+E} & 1 & 1.43 (2.36) & -20.53 (5.58) & 3.61 (3.23) & \textbf{-1.25 (1.86)} & -21.46 (3.77) & 11.80 (2.68) \\
\midrule
\syscell{Proposed}{M}   & 3 & -5.13 (1.98) & \textbf{-22.12 (5.77)} & 10.59 (1.43) & -3.45 (5.78) & -22.62 (5.48) & 10.25 (1.90) \\
\syscell{Proposed}{M+E} & 3 & 2.21 (2.49) & -20.38 (5.37) & 2.75 (4.35) & -1.29 (3.26) & -21.90 (4.23) & 9.37 (3.23) \\
\midrule
\syscell{Proposed}{M}   & 5 & -3.85 (4.78) & -21.50 (5.93) & 9.09 (1.95) & -9.24 (7.76) & -23.47 (4.94) & 9.66 (1.81) \\
\syscell{Proposed}{M+E} & 5 & -2.72 (2.82) & -21.32 (5.99)  & \textbf{0.81 (2.01)} & -6.64 (5.22) & -22.38 (4.07) & 9.30 (2.69) \\
\midrule
\syscell{Proposed}{M}   & 7 & -3.12 (2.65) & -21.74 (6.02) & 8.49 (1.79) & -13.02 (8.65) & -22.74 (5.57) & 9.00 (1.61) \\
\syscell{Proposed}{M+E} & 7 & \textbf{-0.05 (3.23)} & -20.90 (6.00) & 1.35 (2.65) & -13.00 (7.86) & -22.95 (4.49) & \textbf{8.66 (3.80)} \\
\midrule

\syscell{ISM Dataset}{} & 7 & -- & -- & -- & -14.76 (5.92) & -21.86 (5.36) & 10.09 (3.30) \\
\bottomrule 
\end{tabular}
\end{adjustbox}
\vspace{-1em}
\label{tab:evaluation}
\end{table*}

\subsection{Experiment 2: CFG training with combined dataset}
\begin{figure}[H]
  \centering
  \begin{subfigure}{0.85\linewidth}
    \centering
    \includegraphics[width=\linewidth]{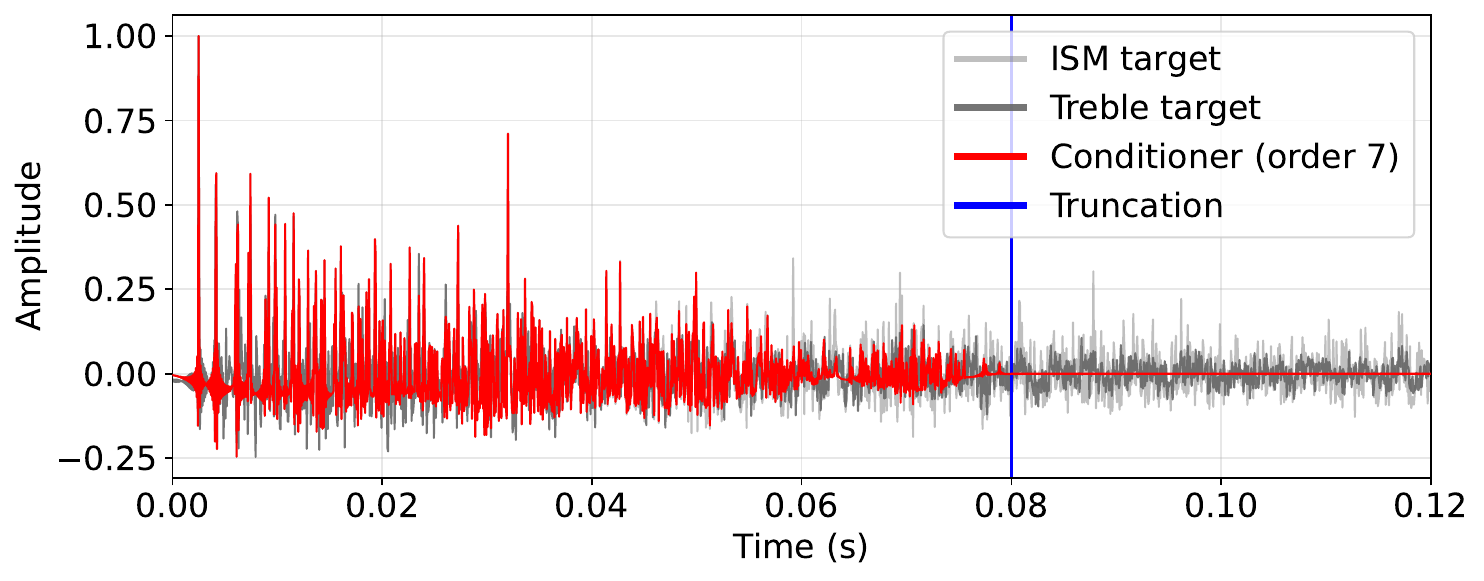}
    \vspace{-5mm}
    \caption{A sample conditioner of order $5$ and target RIRs.}
    \label{fig:conditioner_targets}
  \end{subfigure}


  \begin{subfigure}{0.85\linewidth}
    \centering
    \includegraphics[width=\linewidth]{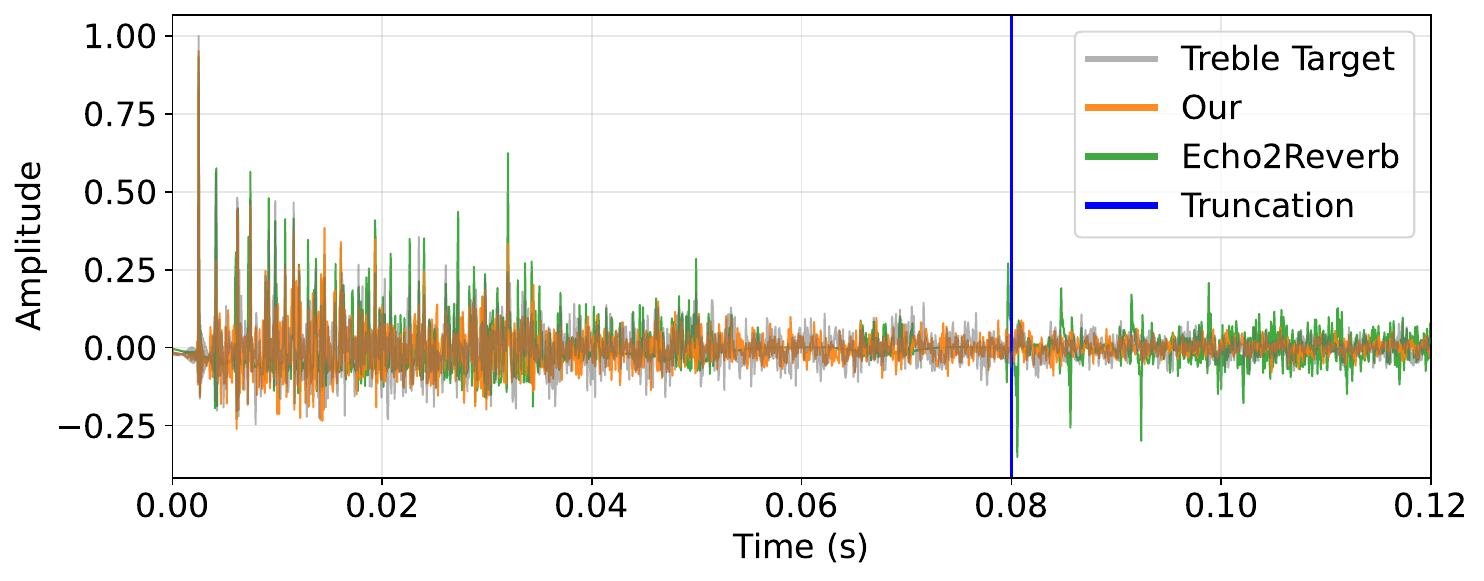}
    \vspace{-5mm}
    \caption{Target and predicted early RIRs.}
    \label{fig:prediction_early}
  \end{subfigure}

  \begin{subfigure}{0.85\linewidth}
    \centering
    \includegraphics[width=\linewidth]{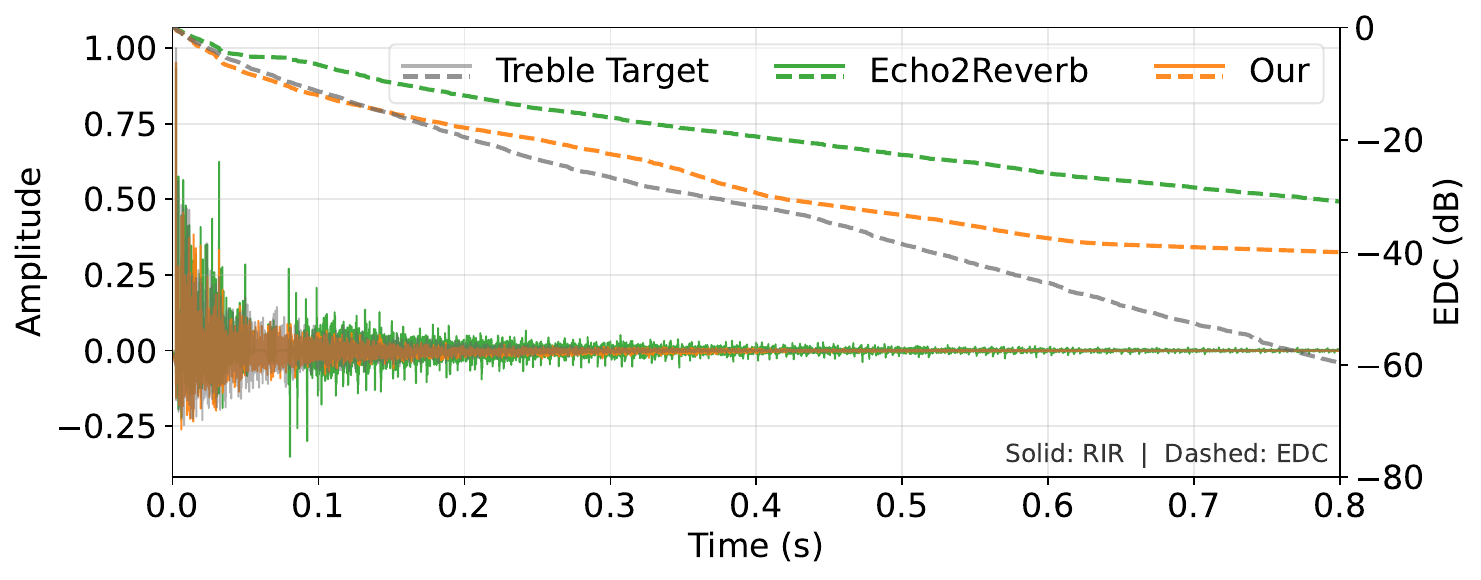}
     \vspace{-5mm}
    \caption{RIRs and EDCs of target and predicted RIRs.}
    \label{fig:predictions_longer}
  \end{subfigure}
        \vspace{-2mm}
      \caption{Example comparisons in the test dataset for Exp.~2. (a) Conditioners of maximum reflection order 5 and target RIRs from both ISM and Treble datasets. The final target of Exp.~2 is always from the Treble dataset. (b) Early RIRs of the target and predictions of our method using the hybrid loss and Echo2Reverb. (c) The RIRs and EDCs for a longer duration.}
  \label{fig:two_vertical}
\end{figure}

In the following, the ISM dataset is also compared with the Treble dataset to demonstrate the intrinsic dataset mismatch between the two datasets, which provides a cross-dataset discrepancy baseline. In this case, the ISM dataset and Treble dataset are used as the prediction and target, respectively. The discrepancy baseline is illustrated as the last row in Tab.~\ref{tab:evaluation} for the input conditioner of maximum reflection order 7, which almost fills the early $80$~ms window, leaving little room for early RIR completion. It is clear that both our method and Echo2Reverb outperform the discrepancy baseline at most configurations for all three metrics, except that our method may have worse EDC error when the conditioner is of order 1. 
Now, the early RER $\textrm{RIR}^{\leq \textrm{80}}_{\textrm{RER}}$ changes substantially with the maximum reflection order for our method. When the orders are $1,3,5$, the early RERs of our method are all better than those in Echo2Reverb. Note that in Exp.~1, this performance advantage is not as obvious since the conditioner and the target are both from the ISM dataset, sharing very close early reflection patterns, as also observed in Fig.~\ref{fig:conditioners}. However, in Exp.~2, the Treble dataset is used as the target. An example comparison of the early RIRs between the ISM and Treble datasets is in Fig.~\ref{fig:conditioner_targets}, indicating that differences are already visible even in the $80$~ms window due to the additional acoustic effects, e.g., reflections and diffractions caused by furniture. Since Echo2Reverb does not improve the $80$~ms window, the advantageous performance of our method is clearly showcased when the target RIR is from a different dataset to the conditioner. However, when the reflection order of the conditioner increases in our method, e.g., at order $7$, the RER values also get worse since there is little room to complete the $80$~ms window. 
Our method and Echo2Reverb are competitive regarding the RMSE errors $\textrm{RIR}_{\textrm{RMSE}}^{>\textrm{80}}$, which is similar to the observation in Exp.~1. But the EDC error $\textrm{EDC}_{\textrm{MAE}}$ of our methods for conditioner orders $5$ and $7$ is clearly lower than Echo2Reverb in terms of mean value and standard deviation. An example of the predictions of our method and Echo2Reverb is shown in Fig.~\ref{fig:prediction_early} and Fig.~\ref{fig:predictions_longer}. The predictions of Echo2Reverb may contain a discontinuity if the $80$~ms window is not filled with early reflections. Since it directly takes the conditioner from the ISM dataset as the early parts of the output, early completion errors are much more severe, which can be confirmed by the RER $\textrm{RIR}^{\leq \textrm{80}}_{\textrm{RER}}$ in Tab.~\ref{tab:evaluation}. To maintain the overall shape of the EDC curve, Echo2Reverb produces spurious large-amplitude pulses after the $80$~ms window, while our method can generate smooth RIR decay regardless of the conditioner's reflection order and duration. Overall, the EDC of our method is closer to the target EDC compared to Echo2Reverb, even though Echo2Reverb is very good at predicting the EDCs, as observed in Exp.~1. 
The observations in Exp.~2 highlight the advantages of our method with the CFG training, that even when the input conditioner is dropped out at a ratio of $0.2$, our method achieves better early completion of RIRs and EDC fitting than Echo2Reverb, which is trained without conditioner dropout. 

\section{Conclusions}


We proposed an $\mathbf{x}$-prediction diffusion model conditioned on simulated early reflections for RIR completion. Evaluated on two datasets, i.e., an ISM dataset and a more realistic Treble dataset, our method matches the baseline when trained on ISM data alone. When trained with CFG on a hybrid dataset (80\% ISM, 20\% Treble) and tested on an unseen subset of the Treble dataset, it outperforms the baseline in early-reflection completion and energy decay curve reconstruction, provided that consistent input conditioners are used. 

Unlike the baseline, which requires complete early reflections within an 80~ms window, our method accommodates lower-order ISM simulations down to first-order reflections. These results demonstrate the viability of generating realistic RIRs from only the direct path and low-order early reflections. Future work includes improving performance at very low reflection orders and accelerating diffusion model inference, both of which are critical for practical deployment.

\bibliographystyle{IEEEtran}
\bibliography{mybib}

\end{document}